\documentclass[twocolumn,superscriptaddress,10pt,pra,showpacs]{revtex4}
\usepackage{amsmath,graphicx,amsfonts,bm,amssymb}
\usepackage{times}
\begin{document}

\title{Physical implementation of topologically decoherence-protected superconducting qubits}

\author{Zheng-Yuan Xue}

\author{Z. D. Wang}
\email{zwang@hkucc.hku.hk}

\affiliation{Department of Physics and Center of Theoretical and
Computational Physics,\\ The University of Hong Kong, Pokfulam Road,
Hong Kong, China}

\author{Shi-Liang Zhu}

\affiliation{Institute for Condensed Matter Physics, School of
Physics and Telecommunication Engineering,\\ South China Normal
University, Guangzhou, China}

\date{\today}

\begin{abstract}
We propose a scenario to physically implement a kind of
topologically decoherence-protected qubit using superconducting
devices coupled to a micro-wave cavity mode with unconventional
geometric operations. It is shown that the two needed interactions
for selective devices, which are required for implementing such
protected qubits, as well as single-qubit gates, can be achieved
by using the external magnetic flux. The easy combination of
individual addressing with the many-device setup proposed in the
system presents a distinct merit in comparison with the
implementation of topologically protected qubits in a trapped-ion
system.
\end{abstract}


\maketitle

Physical implementation of quantum computers has attracted much
attention as they are generally believed to be capable of solving
diverse classes of hard problems. Systems suitable for hardware
implementation of quantum computers should possess certain
properties, such as relatively long coherent time, easy
manipulation and good scalability. With highly developed
fabrication techniques, superconducting quantum interference
devices (SQUIDs) have shown their competence in implementing the
qubits for scalable quantum computation \cite{Makhlin}.
Furthermore, the idea of placing the SQUIDs inside a cavity,
\emph{i.e}., the circuit quantum electrodynamics, has been
illustrated \cite{fluxqubit,chargequbit,yale,zhu} to have several
practical advantages including strong coupling strength, immunity
to noises, and suppression of spontaneous emission.

Decoherence and systematic errors always occur in real quantum
systems and therefore stand in the way of physical implementation of
quantum computers. Generally, larger systems are more sensitive to
decoherence, which makes the scaling of quantum computers a great
experimental challenge. Therefore, how to suppress the infamous
decoherence effects is a main task for scalable quantum computation.
A promising strategy for quantum computation in a fault-tolerant way
is based on the topological idea \cite{kitaev97}, where gate
operations depend only on global features of the control process,
and are therefore largely insensitive to local noises. The Kitaev
model \cite{kitaev97} consists of a class of stabilizer operators
associated with lattice on the torus, which can be put together to
make up a Hamiltonian with local interactions. The four-fold
degenerate ground state of the Hamiltonian coincides with the
protected space of the stabilizer quantum code. Since all the
excited states are separated from the ground states by an energy
gap, the ground states are persistent to local perturbations.
However, it is extremely difficult to directly implement this novel
idea mainly because four-body interactions are notoriously hard to
generate in experiments.

Very recently, Milman \emph{et al}. \cite{Milman07} proposed a
highly symmetrical Hamiltonian
\begin{equation}\label{thxy}
H=-\hbar\chi_x\sum_{i=1}^{M}\left( \sum_{j=1}^{M}\sigma
_{i,j}^{x}\right) ^{2}-\hbar\chi_y\sum_{j=1}^{M}\left(
\sum_{i=1}^{M}\sigma _{i,j}^{y}\right) ^{2},
\end{equation}
which involves only two-body interactions and can be understood as
spins in a 2-dimensional (2D) lattice. Each spin was labeled by its
position in the lattice with $\sigma^{x,y}_{i,j}$ denoting Pauli
matrices of the spin situated at the intersection of the $i$th row
and $j$th column. It has a two-fold degenerate ground state which
may be considered as the two states in a qubit. These states are
generated by non-local operators and thus are naturally protected
against local sources of decoherence. However, comparing with a
size-independent energy gap in the Kitaev Hamiltonian, the energy
gap here is slightly dependent on the number of the involved spins.
A nice feature of the model lies in that it is possible to directly
implement the Hamiltonian (\ref{thxy}) in a trapped-ion system, as
it eliminates a main experimental constrain-the four-body
interactions in the Kitaev model. In addition, the energy gap of
Hamiltonian (\ref{thxy}) remains large as the system size increases,
and thus it provides more efficient protection than the nearest
neighbor model \cite{junction} where energy gap decreases rapidly.

In this paper, we propose an intriguing scenario to implement the
model Hamiltonian (\ref{thxy}) with superconducting devices (serve
as spins in \cite{kitaev97,Milman07}) coupled to a micro-wave cavity
mode with unconventional geometric operations. In trapped-ion
systems, the individual addressing is difficult for large arrays
($N>3$) since the distance between the ions in the center of the
linear array gets smaller \cite{Wineland,Zhu_Duan}. This constrain
makes the linear configuration proposed in Ref. \cite{Milman07}
difficult to be experimentally implemented. Comparing with the above
difficulty, a distinct feature of the present implementation is that
the combination of individual addressing with a many-device setup is
feasible. We will show that, by using the external magnetic flux as
the effective switch tool, the interactions between selective
devices can be introduced. Furthermore, single-qubit gates, which
are required in the global state initialization, can also be
achieved by the same technology. During the implementation of the
operations, the only parameter needs to be tunable is the frequency
of the external magnetic flux. This is a more efficient way of
controlling the system dynamics as it is much easier for
experimental realization. In achieving the implementation of the
needed interactions and given the fact that selective coupling can
be effectively controlled, it is possible to implement topologically
protected qubits  in this architecture. Moreover, the operations to
achieve the above qubit states are based on unconventional geometric
operations, which have been illustrated to have a high fidelity
\cite{zhuunconventional,Leibfried,zhuerror}.

\begin{figure}[tbp]
\centering
\includegraphics[width=8cm]{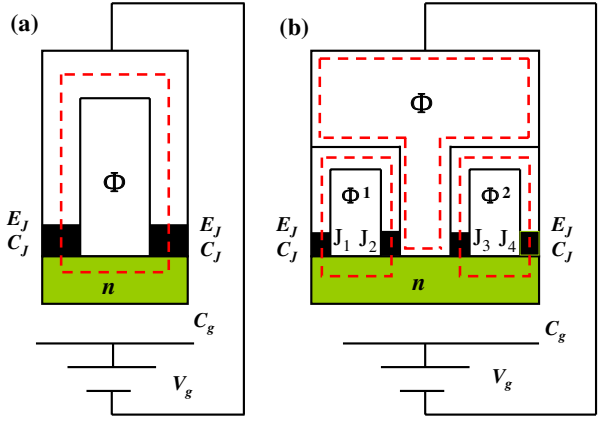}
\caption{(Color online) Schematic illustration of the
superconducting device as the effective spin, the red dash line
denote the integral path of the magnetic flux. (a) A single SQUID as
an effective spin. (b) Device made of two SQUIDs with a common
superconducting charge box. This more flexible design will introduce
more control variables of the effective spin.} \label{qubit}
\end{figure}

A SQUID as a superconducting charge qubit that usually addressed in
the literature is shown in Fig. (\ref{qubit}a) \cite{Makhlin}.
However, it would be clear that this simple configuration can not be
used to acheive the coupling described by Eq. (\ref{thxy}). To
realize the desired Hamiltonian, we design a more flexible device,
as shown in Fig. (\ref{qubit}b), to serve as the effective spin. It
consists of two SQUIDs with a common superconducting charge box,
which has $n$ excess Cooper-pair charges. Each SQUID is formed by
two small identical Josephson junctions (JJs) with the capacitance
$C_{J}$ and Josephson coupling energy $E_J$, pierced by an external
magnetic flux. A control gate voltage $V_g$ is connected to the
system via a gate capacitor $C_g$. The Hamiltonian of the system
reads \cite{Makhlin}
\begin{equation}
\label{h1} H=E_{c}(n-\bar{n})^2-E_J\sum_{l=1}^{4}\cos\varphi^l,
\end{equation}
where $n$ is the number operator of (excess) Cooper-pair charges on
the box, $E_{c}=2e^2/(C_g+4C_J)$  is the charging energy,
$\bar{n}=C_g V_g/2$ is the induced charge  controlled by the gate
voltage $V_g$, and $\varphi^l$ ($l=1, 2, 3, 4$) is the
gauge-invariant phase difference between the two sides of the $l$th
JJ denoted as $J_l$ in Fig. (\ref{qubit}b).  The phase differences
$\varphi^l$ are determined from the flux quantization for three
independent loops, that is, $\varphi^{1}-\varphi^{2}=2\phi^1$,
$\varphi^{2}-\varphi^{3} =2\phi,$ and
$\varphi^{3}-\varphi^{4}=2\phi^2$. Since we here focus on the charge
regime, a convenient basis is formed by the charge states,
parameterized by the number of Cooper pairs $n$ on the box with its
conjugate $\varphi=\sum_{l}\varphi^{l}/4$ by the relation $[\varphi,
n]=\text{i}$. At temperatures much lower than the charging energy
and restricting the gate charge to the range of $\bar{n}\in [0,1]$,
only a pair of adjacent charge states $\{|0\rangle,|1\rangle\}$ on
the island are relevant. Under the condition $\phi^1=\phi^2=0$, the
Hamiltonian (\ref{h1}) is then reduced to
\begin{equation}
\label{h1r} H_s=-E_{ce}\sigma^z-2E_{\Phi}\sigma^x,
\end{equation}
where $E_{ce} =2E_c(1-2\bar{n})$,
$E_{\Phi}=E_J\cos\left(\pi\Phi/\phi_0\right)=E_J\cos\phi$ with
$\phi_0=hc/e$ being the normal flux quantum and
$\phi=\pi\Phi/\phi_0$.

When the device is placed in a cavity, the gauge-invariant phase
difference becomes
$\varphi_m^{'}=\varphi_m-\frac{2\pi}{\phi_0}\int_{l_m} {\bf A}_m
\cdot d {\bf l}_m,$ where  ${\bf A}_m$ is the vector potential in
the same gauge of $\varphi_m$. ${\bf A}_m$ may be divided into two
parts ${\bf A}_m^\prime+{\bf A}_m^\phi $, where the first and
second terms arise from the electromagnetic field of the cavity
normal modes and the external magnetic flux, respectively. For
simplicity, we here assume that the cavity has only a single mode
to play a role. Therefore, we have $\frac{2\pi}{\phi_0}\int_{l_m}
{\bf A}_m\cdot d {\bf l}_m =\frac{2\pi}{\phi_0} \int_{l_m} {\bf
A}^\phi_m\cdot d {\bf l}_m + 2g\left(a+a^{\dagger}\right),$ where
$2g$ is the coupling constant between the junctions and the
cavity, $a$ ($a^\dagger$) is the annihilation (creation) operator
for the cavity mode, with $\omega_c$ being its frequency, the
closed path [the red dash line in Fig (\ref{qubit})] integral of
the ${\bf A}^\phi$ gives rise to the magnetic flux $\Phi$. Then
the Hamiltonian (\ref{h1r}) becomes
\begin{eqnarray}\label{hc1}
H_c=-E_{ce}\sigma_z-2E_J\cos\left[\phi+g\left(a+a^{\dagger}\right)\right]\sigma_x.
\end{eqnarray}

We now show that the required single-qubit gates can be achieved by
the dc magnetic flux. Generally speaking, since the coupling
constant $g$ is very small comparing with $\phi$, we may expand the
Hamiltonian (\ref{hc1}) up to the first order of $g$. In the
rotating frame with respect to
$H_{dc}=-E_{ce}\sigma_z-2E_{\Phi}\sigma_x+\hbar\omega_c\left(a^{\dagger}a+\frac{1}{2}\right),$
the strength of the cavity-SQUID interaction term in Eq. (\ref{hc1})
is proportion to $1/(\omega_c-\omega_q)$ with
$\omega_q=2E_c(1-2\bar{n})/\hbar$, which is an experimentally
controllable parameter via the gate voltage $V_g$. In the present
scheme, we choose $\omega_{q}=0$, which corresponds to the
degeneracy point. Given the facts that $g$ is relatively small and
the frequency difference $\omega_c-\omega_q$ is chosen to satisfy
the large-detuning limit, the cavity mediated interaction can thus
be safely neglected. In other words, the device and the cavity
evolve independently in this case. The external flux is merely used
to address separately the single-qubit rotations, while the
evolution of the SQUID is governed by Hamiltonian (\ref{h1r}).

\begin{figure}[tbp]
\centering
\includegraphics[width=8cm]{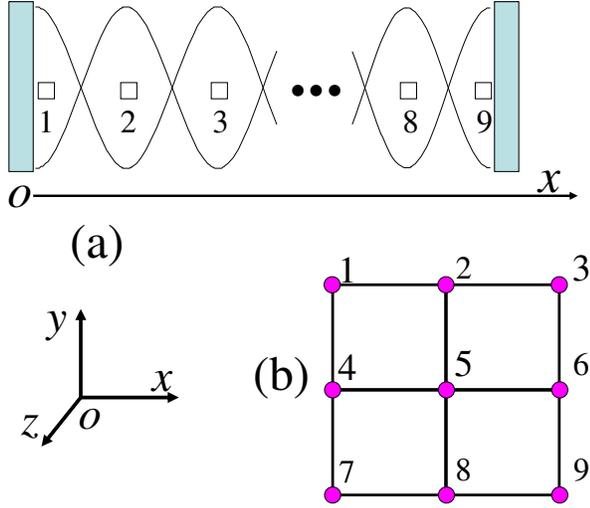}
\caption{(Color online) (a) Schematic superconducting devices
(denoted as rectangles) array constructed along the cavity direction
$x$ with $N=9$. The chosen coordinate is shown in the left-lower
panel. Each device is placed parallel to the $xoz$ plane and at the
antinodes of the single-mode standing-wave cavity. The magnetic
component of the cavity mode is along the $y$ direction, which is
perpendicular to the device loop plane, so that it is the only
contributed component. Devices are placed at the antinodes of the
cavity mode, so that the device-cavity interaction constant ($2g$)
of different devices can be treated as the same constant. (b) An
$N=9$ linear array corresponds to a 2D $3\times3$ array, where the
filled dots with the number represents the devices. For example, if
one wants to generate the interaction in row 1, then devices 1, 2
and 3 are selected to achieve the $J_x^2$ coupling via the virtue
excitation of the cavity mode, while the $J_y^2$ coupling among
devices 2, 5 and 8 is implemented in order to achieve the
interaction in column 2.} \label{array}
\end{figure}

The required interactions between the selected devices may be
induced by using the ac magnetic flux with devices in Fig.
(\ref{qubit}b). If $N$ devices are located within a single-mode
cavity as shown in Fig. (\ref{array}a), to a good approximation, the
whole system can be considered as $N$ two-level systems coupled to a
quantum harmonic oscillator \cite{chargequbit}. The Hamiltonian of
the \emph{N} such devices, placed in a single mode cavity, reads
\cite{zhu}
\begin{equation}  \label{H_single}
H=\sum_{j=1}^N \left[E_{ce} (n_j-\bar{n}_j)^2-E_J\sum_{l }
\cos\varphi_j^{l}\right],
\end{equation}
where we have assumed $E_{ce,j}=E_{ce}$ and $E_{J,j}^{l}=E_J$. Here
$\phi^1$ and $\phi^2$ are the dc magnetic flux, and we have
neglected the cavity-mediated interaction parts in these two SQUID
loops considering that they can be made to be very small comparing
with the inter-SQUID loop, or we can just put the magnetic shield to
exclude the influence of the cavity mode magnetic flux in these two
loops. The phase relation in the inter-SQUID loop is now modified as
$\varphi_j^{2}-\varphi_j^{3} =2\phi_j+2g_j(a+a^{\dagger})=2\omega
t+2g(a+a^{\dagger})$. Assuming all the devices work in their
degeneracy points, under Lamb-Dicke (LD) limit and rotating-wave
approximation as well as in the interaction picture with respect to
$H_0 =\hbar\omega_c(a^\dagger a+\frac{1}{2})$, the cavity mediated
interaction can be described by the Hamiltonian
\begin{eqnarray}
\label{hint2}
H_{int}=\text{i}\hbar\beta\sum_{j=1}^N\sigma_j^{\dagger}\left(\eta
a^{\dagger}e^{\text{i}\delta t} -\xi ae^{-\text{i}\delta
t}\right)+\text{H.c}.,
\end{eqnarray}
where $gE_J/2\hbar=\beta\ll\delta=\omega_c-\omega\ll\omega$,
$\sigma^{\pm}=\frac{1}{2}(\sigma^x \pm \text{i}\sigma^y)$,
$\eta=e^{-\text{i}\phi_-}+e^{-\text{i}\left(2\phi_+-\phi_-\right)}$,
$\xi=e^{-\text{i}\phi_-}+e^{\text{i}\left(2\phi_++\phi_-\right)}$,
and $\phi_{\pm}=\frac{1}{2}\left(\phi^1\pm\phi^2\right)$. If
$\phi_-=0$ and $\phi_+=k\pi$, then Eq. (\ref{hint2}) reduces to
\begin{eqnarray}
\label{hintxx}
H_{int}^{x}=2\text{i}\hbar\beta\left(a^{\dagger}e^{\text{i}\delta
t}-ae^{-\text{i}\delta t}\right)J_x,
\end{eqnarray}
where $J_{x, y, z}=\sum_{j=1}^N\sigma_j^{x, y, z}$. The
corresponding effective Hamiltonian is given by \cite{james,sm}
\begin{eqnarray}
\label{hx} H_x=-\hbar\chi_x J_x^2,
\end{eqnarray}
where $\chi_x=4\beta^2/\delta$. If $\phi_-=\pi/2$ and
$\phi_+=k\pi+\pi/2$, then Eq. (\ref{hint2}) reduces to
\begin{eqnarray}
\label{hintyy}
H_{int}^{y}=2\text{i}\hbar\beta\left(a^{\dagger}e^{\text{i}\delta
t}-ae^{-\text{i}\delta t}\right)J_y,
\end{eqnarray}
with the effective Hamiltonian being given by
\begin{eqnarray}
\label{hy} H_{y}=-\hbar\chi_y J_y^2,
\end{eqnarray}
where $\chi_y=\chi_x$. Actually the configuration described in Fig.
(\ref{qubit}a) is exactly equivalent to the device in Fig.
(\ref{qubit}b) in the specific case of $\phi^1=\phi^2\equiv0$. So it
is now seen that the required Hamiltonian (\ref{thxy}) is unable to
be achieved with the one SQUID in Fig. (\ref{qubit}a) since only the
interaction $H_x$ could be realized.

It is clear that the specific choice of the dc magnetic flux can
lead to the designated interaction type of the selected device.
The cavity-device coupling and decoupling can be controlled by
selecting the external magnetic flux to be dc or ac, not simply by
changing the parameters of the device or the cavity. Since the
external magnetic flux can be effectively controlled, the
cavity-device interaction can be implement in the selected
devices. In addition, all the devices are always stay in their
degeneracy points in this implementation, where they possess long
coherence time and minimal charge noises. Interestingly, the
interactions in Eqs. (\ref{hx}) and (\ref{hy}), which correspond
to the row and column interactions in Eq. (\ref{thxy}), are
insensitive to the thermal state of the cavity mode. It is notable
that the evolutions governed by the Hamiltonians (\ref{hx}) and
(\ref{hy}) may also be considered as unconventional geometric
operations \cite{zhuunconventional}, which are robust against
random operation errors \cite{zhuerror}, and thus have been
essentially used in quantum information processing
\cite{zhu,Zhu_Duan,zhuunconventional,Leibfried,sm,uqc}.

With the two interactions in Eqs. (\ref{hx}) and (\ref{hy}), and
given the fact that they can be implemented on selected devices, it
is straight forward to implement the highly symmetrical Hamiltonian
(\ref{thxy}). The interaction exists only between the devices
selected at one time, thus the spatial pattern of the physical array
is not necessarily related to the 2D configuration directly. A
possible configuration rendering the implementation of the
Hamiltonian easier in existing systems consists of using a linear
array of devices \cite{Milman07}. To implement a topologically
protected logical qubit consists of $M\times M$ physical devices in
2D lattice, we can use a 1D linear array with $N=M^2$ devices, where
the $i$th row and $j$th column in the 2D lattice correspond to the
devices $(i-1)\times M+1, (i-1)\times M+2, ..., i\times M$ and $j,
j+M, ..., j+(M-1)\times M$ in the 1D array, respectively. A small
$3\times3$ 2D array is illustrated in Fig (\ref{array}b) as a
example. After scaling the interactions of each row and column, the
resulting Hamiltonian is the sum of the terms in Eq. (\ref{thxy}).
The coupling between rows and columns may be avoided \cite{Milman07}
by applying operations that alternate between rows and columns
provided that each operation time $\tau$ satisfies
$\tau\chi_{x;y}\ll1$.

Topologically protected qubits usually consist of many physical
qubits and thus seem to be resource-consuming strategy. In fact, a
small finite array, \emph{e.g}., $M=3$ for Hamiltonian (\ref{thxy})
\cite{Milman07} and $M=5$ for nearest neighbor model
\cite{junction}, are likely sufficient to provide good protection
from the noise by suppressing its effect to be many orders of
magnitude lower, which could meet most practical demands.

One experimentally feasible procedure to generate a topologically
protected qubit state is as follows. (i) Applying a large effective
field along the $x$ direction (putting $\bar{n}=1/2$ and $\phi=0$ or
$\phi=\pi$) when devices are decoupled from the cavity, we get
$H_1=\pm E_J\sum\sigma_{i,j}^x$. Then the initial state is logical
$|0\rangle=\prod |0\rangle_{ij}$ or logical $|1\rangle=\prod
|1\rangle_{ij}$ depending on the overall sign in the Hamiltonian
$H_1$. (ii) Adiabatically switching off $H_1$ and then switching on
the Hamiltonian (\ref{thxy}). After performing these two steps, the
two initial unprotected logical states $|0\rangle$ and $|1\rangle$
evolve into one of the protected ground states of the model
Hamiltonian (\ref{thxy}).

We now briefly address the experimental feasibility in implementing
our scheme. Suppose that the quality factor of the superconducting
cavity is $Q=1\times10^{6}$. For the cavity mode with
$\omega_c/2\pi=50$ GHz \cite{qedreview}, the cavity decay time is
$\tau_c\approx3.2$ $\mu$s. With a moderate vacuum Rabi frequency
$\Omega\sim10$ MHz and the lifetime of the SQUID $\gamma=2$ $\mu$s,
the strong-coupling limit can be readily reached \cite{yale}.
Meanwhile, $2g=10^{-2}$ \cite{zhu} to ensure the LD limit. With
$E_J=40$ $\mu$eV \cite{zhu}, $\beta/2\pi\approx48$ MHz. To maintain
the large-detuning condition,  we can choose $\delta\sim10\beta$,
which in turn satisfies readily the requirement of
$\delta\ll\omega\sim\omega_c$. A typical operation time for
large-detuning interaction is $t\approx10$ ns \cite{zhu}, which
ensures that thousands of operations are possible \cite{yale} since
the coherence time of the SQUID and cavity mode are both in the
order of $\mu$s. When all devices are located at the antinodes of
the microwave, the distance of the two neighbor devices is a half of
the microwave wavelength. For the chosen cavity mode ($\lambda_c=6$
mm), 10 devices may be constructed along the cavity direction
\cite{qedreview}. Typically, a SQUID loop ($1\times1$ $\mu$m)
consists of two small identical JJs of the size $0.1\times0.1$
$\mu$m, and the inter-SQUID loop can be fabricated in the size of
$4\times4$ $\mu$m. Therefore, the distance of the two devices is
about $10^3$ times of the device size, so that their mutual
induction may be safely neglected. Moreover, it is notable that
other kinds of 4-junction qubits have already fabricated and used in
the experiments, \emph{e.g}., in \cite{mooij}. Thus, the current
design of qubit impose no extra requirements on the current
technology of the qubit fabrication process. Finally, we roughly
estimate the effect of the nonuniformity in device parameters,
$E_{ce,j}$and $E_j^l$, to the operations in our implementation. As
different devices can be addressed and controlled individually, it
is easy to tune each device in its degeneracy point via its gate
voltage separately, such that the nonuniformity of $E_{ce}$ among
different devices is not of significant importance. The
nonuniformity of the Josephson couplings with a deviation of
$\epsilon E_j$ will cause a minor change of the operation infidelity
in the order of $\exp(-\Delta/\epsilon\chi_x)$,  which is about
$1\%$ and thus can be neglected for the reasonable experimental
parameters $\epsilon\sim 20\%$ and $\delta\sim 10 \beta$, where
$\Delta\sim \chi_x$ is the energy gap of topological phase.

\bigskip

We thank L. B. Shao and G. Chen for fruitful discussions. This work
was supported by the RGC of Hong Kong under Grant Nos. HKU7045/05P
and HKU7049/07P, the URC fund of HKU, the NSFC under Grant Nos.
10429401 and 10674049, NCET and the State Key Program for Basic
Research of China (Nos. 2006CB921800 and 2007CB925204).

\end{document}